\documentclass{ifacconf}

\usepackage[usenames, dvipsnames]{color}
\usepackage{amssymb}
\usepackage{mathtools}
\usepackage{enumerate}
\usepackage{amsmath}
\usepackage{array}
\usepackage{algorithm}
\usepackage[noend]{algpseudocode}
\usepackage[noadjust]{cite}
\usepackage[normalem]{ulem}
\useunder{\uline}{\ul}{}
\usepackage{pdfpages}
\newtheorem{theorem}{Theorem}
\newtheorem{remark}{Remark}
\newtheorem{definition}{Definition}

\usepackage{graphicx}
\graphicspath{{Images/}}
\DeclareGraphicsExtensions{.pdf,.emf,.png}
\usepackage{natbib}        
\begin{document}
\begin{frontmatter}

\title{Robust Fault Diagnosis by Optimal Input Design for Self-sensing Systems\thanksref{footnoteinfo}} 

\thanks[footnoteinfo]{This work was supported by Oc\'e Technologies B.V.}

\author[First]{Dhruv Khandelwal} 
\author[First]{Siep Weiland} 
\author[Second]{Amol Khalate}

\address[First]{Department of Electrical Engineering, Eindhoven University of Technology, 
   5600MB Eindhoven, Netherlands (e-mail: D.Khandelwal@tue.nl, S.Weiland@tue.nl).}
\address[Second]{Oc\'e Technologies B.V., 
   5914 CA Venlo, Netherlands, (e-mail: amol.khalate@oce.com)}

\begin{abstract}                
This paper presents a methodology for model based robust fault diagnosis and a methodology for input design to obtain optimal diagnosis of faults. The proposed algorithm is suitable for real time implementation. Issues of robustness are addressed for the input design and fault diagnosis methodologies. The proposed technique allows robust fault diagnosis under suitable conditions on the system uncertainty. The designed input and fault diagnosis techniques are illustrated by numerical simulation.
\end{abstract}


\end{frontmatter}


\section{Introduction}
\label{sec:Intro}

In fault diagnosis, one usually distinguishes between methods that do or do not involve the use of auxiliary input signals to distinguish faults. These are referred to as active and passive fault diagnosis respectively (\cite{vsimandl2009active}). Passive fault diagnosis techniques utilize measurements obtained from the system during routine operation to detect and diagnose faults. The input applied to the system is not considered as a degree of freedom in these methods. Typically, the available input-output data is compared to data inferred from existing models to detect faults.  This technique, however, may not be sufficient to detect all faults. Active fault diagnosis utilizes an auxiliary input test signal solely for the purpose of detection of faults. A well designed input can significantly increase the probability of detecting faults from measured data of the system. These methods were first introduced in \cite{zhang1989auxiliary} and are typically implemented with the aim of taking remedial actions during the operation of the system. Moreover, the auxiliary signal is typically required to have minimal impact on the usable output of the system. The book of \cite{patton2013issues} contains a thorough overview of various fault detection, isolation and diagnosis mechanisms, including active and passive fault diagnosis, developed over the years.%

Several methods and algorithms for active fault diagnosis have been proposed in recent years. In \cite{campbell2004auxiliary}, a methodology for design of an auxiliary input signal is proposed that ensures the detectability of faults in linear systems with deterministic uncertainties. The input design method is restricted to two systems - one nominal and one faulty. The optimal input is defined as the input with minimal norm that ensures the detectability of faults. This method has since been extended to systems with a priori information of initial conditions (\cite{nikoukhah2006auxiliary}), discrete time models (\cite{ashari2011auxiliary}), and non-linear systems (\cite{andjelkovic2008active}). \cite{kim2013optimum} propose a method of input design for fault diagnosis by adopting a statistical approach to distinguish between nominal and faulty systems. Unlike \cite{campbell2004auxiliary}, this method can be used for multiple faulty models, and models the uncertainties using stochastic differential equations. An active fault diagnosis method for nonlinear systems with probabilistic uncertainties is presented in \cite{mesbah2014active}.

The paper of \cite{olaru2010positive} proposes a control scheme for multiple sensor systems which, in real time, chooses a sensor to ensure fault tolerant closed loop operation of the system. While this work focuses on sensor faults, \cite{stoican2011adaptation} adapt the theory for a generalized fault detection and diagnosis problem. This method uses the concept of minimal robust positive invariant sets to characterize the nominal and faulty systems and uses set membership techniques to identify faults. These studies do not deal with the problem of optimal input design for fault detection. The method presented in \cite{scott2014input} involves using zonotopes to characterize a set of inputs that guarantee the diagnosis of faults. Subsequently, a norm-minimizing input is computed over this set of `seperating inputs' without the need of computing the set. This method has been demonstrated to be computationally more efficient in comparison with \cite{nikoukhah1998guaranteed}.


In this paper, a method of active fault diagnosis for LTI systems with additive perturbations is proposed. While the proposed method remains applicable for all systems belonging to the aforementioned class, it is of particular interest for self-sensing systems, such as piezo-electric actuated systems. In terms of data-acquisition, self-sensing systems pose a unique challenge: the system is either in an (input) excitation mode or in a (output) measurement mode. We propose a fault diagnosis algorithm that is computationally efficient and suitable for applications with stringent real time constraints, as prevalent in high-tech systems and safety-critical applications. Additionally, a novel method of optimal and robust input design is presented. The input is designed offline. The method proposed here leads to system responses that allow a guaranteed correct diagnosis of faults in a prescribed (finite) time window and in the face of uncertainties in the system. We establish well quantified conditions on the size of the prescribed set of uncertainties that permit guarantees on correct fault diagnosis. The proposed methodologies are illustrated by a numerical simulation.

The paper is organized as follows. Section \ref{sec:problemFormulation} introduces the setting of the fault diagnosis experiment, and provides the problem formulation. Some preliminaries and mathematical concepts are introduced in Section \ref{sec:detCaseMath}. In Section \ref{sec:IDFramework}, we present the proposed methodology for input design for the exact case. The proposed strategy and sufficient conditions for guaranteed robust fault diagnosis is presented in Section \ref{sec:robust}.  This leads to the extension of input design for the robust case. In Section \ref{sec:example}, the proposed methodologies are demonstrated by an illustrative example. Finally, the conclusions and recommendations are presented in Section \ref{sec:conclusions}. Specific technical material has been collected in the Appendix.
%
%
\section{ Problem Description}
\label{sec:problemFormulation}
%
%
	\subsection{Experiment Setting}
	Applications of active fault diagnosis often involve stringent real time constraints. Typically, one would like to diagnose a fault as quickly as possible in order to take remedial actions or in order to ensure safe operation of the system. The time constraint typically amounts to assuming that a finite amount of time is given to detect and identify faults. In this work, we make use of an \emph{experiment} conducted on the physical system for the purpose of fault diagnosis. The experiment time is finite and is assumed to be divided in an \emph{excitation} window (of length, say, $T_->0$) and a disjoint \emph{measurement} window (of length, say, $T_+>0$). During the excitation window, the system is excited with a pre-designed input. During the measurement window, the transient response of the system is observed and used to diagnose faults. This distinction becomes especially relevant in the context of self-sensing systems.
	
	It is assumed that the system, and each of the possible faults that need to be identified have been modelled as LTI systems. In many applications, the particular model of a faulty system may be uncertain to a prescribed degree. To take this dependence into account, each model is assumed to be subject to norm-bounded additive perturbations. The norm bounds on the additive perturbations are assumed to be known.
	
	
	\subsection{Problem Formulation}
	
	Introduce stable LTI systems $G_0, \dots, G_n$, where $G_0$ is nominal and $G_1, \dots, G_n$ are $n$ possible faults.  Each system $G_i$ is assumed to be an element $G_{i,\Delta} = G_i + \Delta_i$ in a set of uncertain systems 
	\begin{equation}
		\mathcal{G}_i \coloneqq \{G_i + \Delta_i | \Delta_i \in \mathbf{\Delta}_i \},
	\end{equation}
	where $\mathbf{\Delta}_i$ is a set of prescribed norm-bounded stable LTI perturbations that may act on the system $G_i$. In addition, suppose there exists a true unknown system $G_u$, which is subject to unknown additive perturbations, in the sense that $G_u$ belongs to $G_{u,\Delta} := G_u + \Delta_u$, with $\Delta_u \in \mathbf{\Delta}_u$, a set of LTI perturbations with unknown norm-bound. Additionally, we assume without loss of generality that $0 \in \mathbf{\Delta}_i \ \forall i \in [0,n]$. Define the sets $\mathcal{N}_G := \{0,\dots, n\}$ and $\mathcal{S}_G := \{ G_0, \dots, G_n\}$.
	
	 The research problem is then stated as follows. Design an input $u^*$ such that the data $(u^*,y)$, with support on $[-T_-,T_+]$, inferred from $G_{u,\Delta}$, allows to uniquely diagnose $i^* \in \mathcal{N}_G$ such that
	 \begin{equation}
	 	G_{u,\Delta} \in \mathcal{G}_{i^*}.
	 \end{equation}
	 We refer to this as the Optimal Input Design Problem (\emph{OIDP}). Moreover, determine an algorithm to find $i^*$ from the data $(u^*,y)$. We refer to this as the Robust Fault Diagnosis Problem (\emph{RFDP}). Note that the OIDP is closely related to the RFDP - the input design method must be in line with the fault diagnosis technique used. Thus, one may consider the OIDP and the RFDP as two aspects of the same problem. While the OIDP leads to an offline computation of the optimal input, the RFDP involves using the pre-computed input to diagnose faults online.
	 
	 Note that the set of LTI perturbations $\mathbf{\Delta}_u$ on the unknown true system $G_{u,\Delta}$ is unknown. In the following sections, sufficient conditions of the size of $\mathbf{\Delta}_u$ will be established, that permit guaranteed fault diagnosis for the set of systems $\mathcal{S}_G$.
	

\section{Mathematical preliminaries and notation}
\label{sec:detCaseMath}
The unique constraint posed by self-sensing systems fits naturally in a Hankel-like framework. Let ${T_-, T_+ > 0}$ be lengths of the excitation and measurement windows, respectively. Define $[-T_-,0)$ and $[0,T_+]$ as `past' and `future' intervals. Define
	\begin{multline}
		\mathcal{U} := \{ {u:[-T_-,T_+] \rightarrow \mathbb{R}} \; | \; {\sum_{k \in [-T_-,0)} |u(k)|^2 = 1}, \\
		\quad {u(k)=0 \; \forall k \in [0,T_+]}\},
	\end{multline}
	i.e. $\mathcal{U}$ is the set of inputs that are normalized on the past interval $[-T_-,0)$, and vanish on future interval $[0,T_+]$. Let $\mathcal{Y}$ be the space of measurement signals defined as $\mathcal{Y} := \{ y:[0,T_+] \rightarrow \mathbb{R}\}$.
 	For any $i \in \mathcal{N}_G$, consider the following representations for system $G_i \in \mathcal{S}_G$:
	
	\begin{enumerate}
	\item
	An input-state-output representation $G_i^{iso}$:
	\begin{align}
			x_i(k+1) &= A_i x_i(k) + B_i u(k) \label{eq:ssrep}\\\nonumber
			y_i(k) &= C_i x(k) + D_i u(k),
		\quad x_i(-T_-) = 0,
	\end{align}
	\item
	A transfer function representation $G_i(z)$. In the sequel, we will drop the argument $z$. However, this is unlikely to cause any confusion, since the other representations are easily distinguishable.
	\item
	The output nulling state-space representation $G_i^{on}$:
	\begin{align}
	\label{eq:ONrep}
			x_i(k+1) &= \mathcal{A}_i x_i(k) + \mathcal{B}_i w_i(k) \\
			v_i(k) &= \mathcal{C}_i x_i(k) + \mathcal{D}_i w_i(k), 
		\quad i \in \mathcal{N}_G, \nonumber
	\end{align}
	The output signal $v_i(k)$ of the output nulling representation is interpreted as a residual signal. By definition, $v(k) = 0$ for all $k$ if and only if the input-output pair $w_i(k) = \mathrm{col}(u(k),y_i(k))$ is compatible with the model $G_i^{on}$ for some state trajectory $x_i$. If $v_i \ne 0$ for all possible state trajectories $x_i$, the pair $(u,y_i)$ is not compatible with $G_i^{on}$. This idea plays a central role in our fault diagnosis methodology.
	It must be pointed out that, while the state vectors used in \eqref{eq:ssrep} and \eqref{eq:ONrep} are not identical in general, it can be shown that for every system of the form \eqref{eq:ssrep}, there exists an ouput nulling representation of the form \eqref{eq:ONrep} with identical state vectors.
	
	\end{enumerate}
	Define the Hankel operator $\mathcal{H}_i: \mathcal{U} \rightarrow \mathcal{Y}$ that admits the relation
	\begin{equation}
			y_i = \mathcal{H}_i u = \sum_{l=-T_-}^{-1} g_i(k-l) u(l), \, k \in [0,T_+], 
	\label{eq:convolution}%
	\end{equation}
	where, $g_i$ is the inverse z-transform of $G_i$.
	The corresponding operator norm, also known as the Hankel norm on finite horizon, is defined as
	\begin{equation}
	\label{eq:Hankelnorm}
		\|H_i\| = \|G_i\|_H := \sup_{u \in \mathcal{U}} \|y_i\|_2,
	\end{equation}
	where $y_i$ is defined in \eqref{eq:convolution}. This norm indicates the largest output gain in the measurement window $[0,T_+]$ with respect to input applied in the experiment window $[-T_-,0)$.
	
	Additionally, the $\ell_2$ induced norm on the measurement window $[0,T_+]$, is defined as:
	\begin{equation}
	\label{eq:l2norm}
	\|G_i\|_{[0,T_+]} = \underset{\left \| u  \right \|_{[0,T_+] \neq 0}}{\sup} \frac{\sqrt{\sum_{k=0}^{T_+} |y_i(k)|^2}}{\sqrt{\sum_{k=0}^{T_+} |u(k)|^2}}.
	\end{equation}
	The $\ell_2$ induced norm in \eqref{eq:l2norm} measures the maximal gain of the system $G_i$, with input $u$ and output $y_i$, both defined on the measurement window $[0,T_+]$.

	\begin{remark}
	 \label{rem:normalization}
	In the coming sections, we will work with output-nulling representations that are normalized with respect to either the Hankel norm or the induced $\ell_2$ norm. See Appendix \ref{app:normalizeON} for details.
	 \end{remark}


\section{Input Design Framework}
\label{sec:IDFramework}
In this section, we introduce the concept of a \textit{discriminatory input} and propose a performance index to measure the optimality of a given input for the purpose of fault diagnosis. This leads to the design of an optimal discriminatory input. The performance measure is also used to determine the feasibility of the diagnosis problem. This section addresses the problem of input design for the exact case, i.e., we assume that $\mathbf{\Delta}_i = 0 \ \forall i \in \mathcal{N}_G$. In Section \ref{sec:robust}, this framework will be extended to the uncertain case.
	
	\subsection{Discriminatory Input}
	In this paper, the fault diagnosis experiment refers to the excitation of the unknown system $G_u$ with a pre-designed input signal ${u:[-T_-,0) \rightarrow \mathbb{R}}$ defined on the experiment window. This experiment returns an output ${y_u:[0,T_+]\rightarrow \mathbb{R}}$ compatible with system $G_u$ and with input ${u:[0,T_+] \rightarrow \mathbb{R}}$ set to 0. The output $y_u$ is discarded and assumed undefined on the experiment window $[-T_-,0)$. Define ${w:=\textrm{col}(u,y_u)}$, defined on the measurement window $[0,T_+]$ only.
	
	To determine compatibility of signal $w$ and a chosen system $G_i^{on}$, one must find the state trajectory $x_i$ that minimizes the residual $v_i$ in \eqref{eq:ONrep}. Due to the Hankel-like setting of the fault diagnosis experiment, it is possible to translate the problem of finding such a state trajectory $x_i$ to the problem of finding a state vector $x_i(0)$, such that the resulting residual $v_i$ in \eqref{eq:ONrep} in minimized. The solution to this problem, in the general case and for the specified Hankel-like setting, is given in Appendix \ref{app:initState}. The optimal state is denoted by $x_{i,0}^*$. Hence, we excite $G_i^{on}$ with trajectory $w$ and initial condition $x_{i,0}^*$, and obtain the minimal residual $v_i$, as in \eqref{eq:ONrep}. A discriminatory input can now be defined as follows.

	\begin{definition}
	A non-zero input $u \in \mathcal{U}$ is discriminatory for systems $G_i$ if $\{v_i = v_j = 0 \} \Rightarrow \{i = j\}$, with $v_i,v_j$ as defined in \eqref{eq:ONrep}, $j \in \mathcal{N}_G$, and the initial conditions chosen to be $x_i^*,x_j^*$ respectively.
	\end{definition}
	
	So, a non-zero input is discriminatory for $G_i$ if it uniquely identifies $G_i$ on the basis of a zero residual $v_i$ of $G_i^{on}$. We define $\mathcal{U}_{\mathrm{disc}} := \{ u \in \mathcal{U} | \; u \text{ is discriminatory for } \mathcal{S}_G \}$ as the set of all discriminatory inputs for the systems in $\mathcal{S}_G$.
%
%
	\subsection{Optimal Discriminatory Input with respect to $G_i$}
	\label{sec:IDexact}
\begin{figure} [!t]
				\centering
				\includegraphics[scale = 0.35]{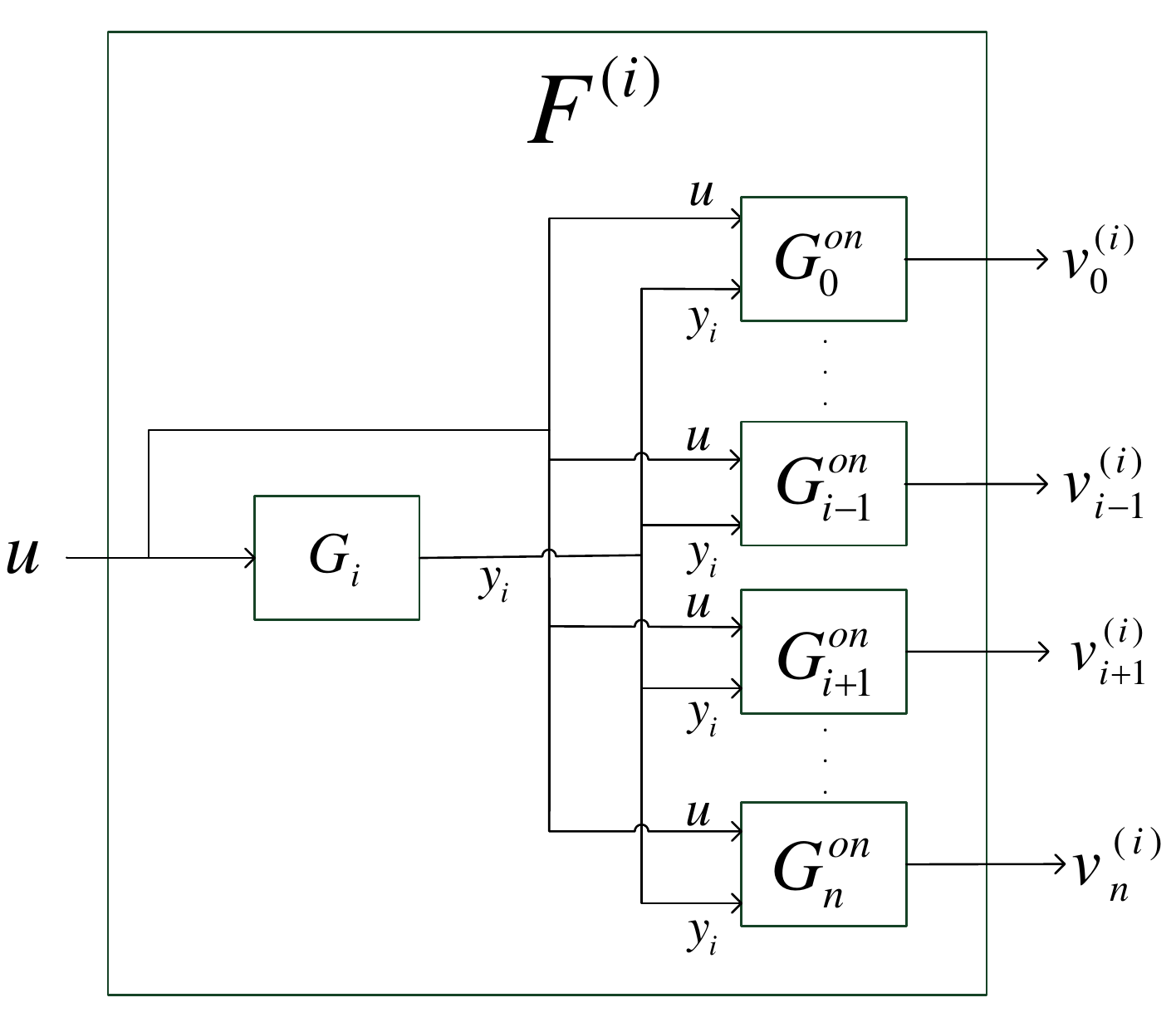}
				\caption{System interconnection $F^{(i)}$ for some fixed $i$.}
				\label{fig:inputDesignFramework1}
			\end{figure}

	Consider system $G_i$ for some fixed $i \in \mathcal{N}_G$. Define $\mathcal{N}^{\bar{(i)}}_G := \{j \in \mathcal{N}_G| j \neq i\}$. Consider a parallel connection $F^{(i)}$ of systems
	\begin{equation}
		F_j^{(i)} := G_j^{on} \begin{pmatrix}
I\\
G_i
\end{pmatrix}, \; \forall j\in \mathcal{N}^{\bar{(i)}}_G,
	\end{equation}
	shown schematically in Fig. \ref{fig:inputDesignFramework1}. Each system interconnection $F_j^{(i)}$ is normalized with respect to the finite horizon Hankel norm (see Appendix \ref{app:normalizeON}). 
Let $\mathcal{H}_j^{(i)}$ be the finite time Hankel operator associated with $F_j^{(i)}$, and let $\mathcal{H}^{(i)}$ be the Hankel operator associated with the system $F^{(i)}$. Let $V^{(i)}$ be the stacked output $\textrm{col}(v_0^{(i)}, \dots , v_{n-1}^{(i)})$ defined on $[0,T_+]$.
			Let,
			\begin{eqnarray}
			\label{eq:ssFij}
				\zeta^{(i)} (k+1) &=& A^{(i)} \zeta^{(i)}(k) + B^{(i)} u(k); \quad \zeta^{(i)} (-T_-) = 0, \nonumber \\
				v_j^{(i)}(k) &=& C_j^{(i)} \zeta^{(i)} (k) + D_j^{(i)} u(k),
			\end{eqnarray}
			 be a state space representation of $F_j^{(i)}$ and let $P^{(i)}$ and $Q_j^{(i)}$ be the corresponding reachability and observability grammian on the finite intervals $[-T_-,0)$ and $[0,T_+]$, respectively, i.e.,
			 \begin{eqnarray}
			 P^{(i)} &:=& \sum_{k=-T_-}^{-1} A^{(i)k} B^{(i)} B^{(i)^{\top}} A^{(i)k^{\top}},\\
			 Q_j^{(i)} &:=& \sum_{k=0}^{T_+-1} A^{(i)k^{\top}} C_j^{(i)^{\top}} C_j^{(i)} A^{(i)k}.
			 \end{eqnarray}
			 Define the performance index $\gamma^{(i)}:\mathcal{U}\rightarrow \mathbb{R}$ as the smallest norm of the $(n-1)$ residuals generated by $F^{(i)}$ when excited with input $u$, i.e.
			 \begin{equation}
			 \gamma^{(i)}(u) := \underset{j \in \mathcal{N}^{(\bar{i})}_G}{\min} \|H_j^{(i)}u\|^2 = \underset{j \in \mathcal{N}^{(\bar{i})}_G}{\min} \|v_j^{(i)}\|_{[0,T_+]}^2, \label{eq:ID_Fi_lb1}
			\end{equation}
			where $v_j^{(i)}$ is the residual defined on $[0,T_+]$, of \eqref{eq:ssFij} with input $u \in \mathcal{U}$ defined on $[-T_-,0)$ and $u=0$ on $[0,T_+]$.
			\begin{definition}
			An input $u^{(i)}$ is \textit{optimally discriminatory with respect to $G_i$} if it maximizes the performance index  $\gamma^{(i)}(u)$ over the set $\mathcal{U}$.
			\end{definition}
			Note that with ${u \in \mathcal{U}}$, at time $k=0$, the state of \eqref{eq:ssFij} is driven to $\zeta^{(i)}(0)$ that satisfies
			\begin{equation}
				\left( \zeta^{(i)}(0) \right)^\top \left( P^{(i)} \right)^{-1} \zeta^{(i)}(0) = 1,
			\end{equation}
			and the norm of the corresponding residual can be expressed as
			\begin{equation}
				\|v_j^{(i)}\|_{[0,T_+]}^2 = \left(\zeta^{(i)}(0) \right)^\top Q^{(i)}_j \zeta^{(i)}(0).
			\end{equation}
			Define the reachability matrix of the system $F^{(i)}$ as 
			\begin{equation}
				R^{(i)}:= [B^{(i)} \; A^{(i)}B^{(i)} \; \dots \; (A^{(i)})^{T_- - 1} B^{(i)}].
			\end{equation}
			The optimal discriminatory input with respect to $G_i$ is then characterized in terms of the state $\zeta^{(i)}(0)$ in \eqref{eq:ssFij}, in the following lemma.
			\begin{lem}
				The optimal discriminatory input with respect to $G_i$ is given by
				\begin{equation}
					u^{(i)*} = \left(R^{(i)} \right)^\top \left( P^{(i)} \right)^{-1} \zeta_0^{(i)*},
				\label{eq:ID_opt2}
				\end{equation}
				where $\zeta_0^{(i)*}$ is the solution to the optimization problem
				\begin{align}
			\zeta^{(i)*}_0 &= \underset{\zeta_0}{\arg \max} \; \underset{j \in \mathcal{N}^{(\bar{i})}_G}{\min} (\zeta_0)^\top Q^{(i)}_j \zeta_0, \label{eq:ID_opt}\\
			\intertext{subject to,}
			\multicolumn{2}{c}{\centering{ $(\zeta_0)^\top (P^{(i)})^{-1} \zeta_0=1.$}}
			\end{align}
			\end{lem}
			
			The optimization problem \eqref{eq:ID_opt} is a non-linear optimization problem, often treated in the field of game theory. Several algorithms exist that find a locally optimal solution for this problem (for example, see \cite{bard1988convex}, \cite{vicente1994bilevel} and all references therein). 


	\subsection{Optimal Discriminatory Input with respect to $\mathcal{S}_G$}
			\label{sec:IDdeterministic}		
\begin{figure} [!t]
				\centering
				\includegraphics[scale = 0.35]{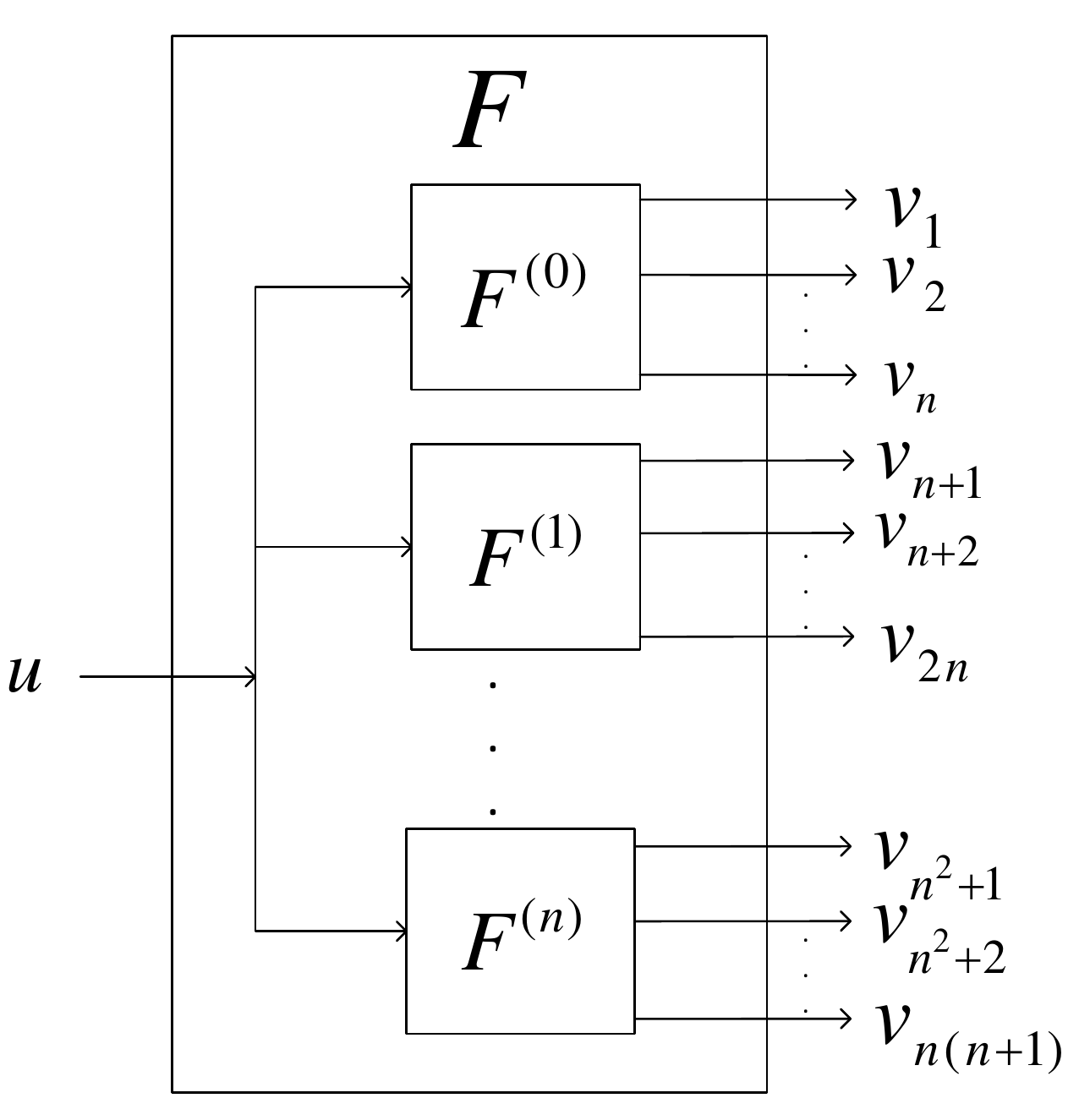}
				\caption{Sytem interconnection $F$.}
				\label{fig:inputDesignFramework2}
			\end{figure}

			Consider a parallel connection $F$ of the systems $F^{(i)}$, for all $i \in \mathcal{N}_G$, as shown in Fig. \ref{fig:inputDesignFramework2}. Each sub-system $F^{(i)}$ consists of a parallel connection of systems $F^{(i)}_j$ as defined in \eqref{eq:ssFij}. $F$ maps input $u \in \mathcal{U}$ to output residual signals $V = \textrm{col}(v_l)$, with $l \in [1,n(n+1)]$. Let $\mathcal{H}_F$ be the Hankel operator associated with this LTI system $F$. Consider a state-space realization of the system $F$ with state vector $\zeta$ such that $\zeta(-T_-) = 0$ and state space matrices $(A,B,C,D)$. This can be inferred from \eqref{eq:ssFij}. As before, let $P$, $Q_l$ be the reachability grammian of the system and the observability grammian for the $l^{th}$ output of the system over the windows $[-T_-,0)$ and $[0,T_+]$ respectively:
			\begin{eqnarray}
			 P &:=& \sum_{k=-T_-}^{-1} A^{k} B B^{\top} A^{k^{\top}},\\
			 Q_l &:=& \sum_{k=0}^{T_+-1} A^{k^{\top}} C_l^{\top} C_l A^{k},
			 \end{eqnarray}
			where $C_l$ is the $l^{th}$ output of $F$. The $i$-independent performance index $\gamma(u)$ is now defined as
			\begin{equation}
				\gamma(u) := \underset{l}{\min} \| \mathcal{H}_Fu\|^2 = \underset{l}{\min} \|v_l\|_{[0,T_+]}^2.
				\label{eq:IDexact_gam_full}
			\end{equation}
				
			\begin{definition}
				An input $u^*$ is called \textit{optimally discriminatory} if it maximizes the performance index $\gamma(u)$ over the set $\mathcal{U}$ for the system $F$.
			\end{definition}
				
			Once again, note that with $u \in \mathcal{U}$, the state $\zeta$ of system $F$ is driven to $\zeta(0)$ that satisfies 
			\begin{equation}
				\zeta(0)^\top P^{-1} \zeta(0) = 1,
			\end{equation}
			and the energy in the corresponding residual can be computed as
			\begin{equation}
				\|v_l\|_{[0,T_+]}^2 = \zeta(0)^\top Q_l \zeta(0).
			\end{equation}
			Define the reachability matrix of system $F$ as 
			\begin{equation}
				R:= [B \ AB \ \dots \ A^{T_- - 1} B].
			\end{equation}
			\begin{theorem}
				The optimal discriminatory input with respect to $\mathcal{S}_G$ is given by:
				\begin{equation}
				u^* = R^\top P^{-1} \zeta_0^*, \label{eq:ustar}
			\end{equation}
			where $\zeta_0^*$ is the solution to the optimization problem
			\begin{equation}
			\label{eq:ID_opt4}
				\begin{split}
					&\zeta_0^* = \underset{\zeta_0}{\arg \max} \; \underset{l}{\min} \; \zeta_0^\top Q_l \zeta_0,\\
					&\text{subject to,}\\
					& \quad \quad \quad \zeta_0^\top P^{-1} \zeta_0 = 1.
				\end{split}
			\end{equation}
			\label{th:optInputSG}
			\end{theorem}
			Theorem \ref{th:optInputSG} provides the solution for OIDP for the \emph{exact case}. In Section \ref{sec:robustID}, we demonstrate that this also corresponds to the solution for OIDP in the robust case, within the framework of the proposed fault diagnosis scheme presented in the next Section.
			
			The proposed performance index can be used to characterize the feasibility of the fault diagnosis problem as follows.
			
			\begin{prop}
				The following statements are equivalent:
				\begin{enumerate}
					\item The fault diagnosis problem is infeasible,
					\item $\mathcal{U}_{\mathrm{disc}} = \phi$,
					\item $\gamma(u^*)=0$, where $u^*$ is the solution obtained from Theorem \ref{th:optInputSG}.
				\end{enumerate}
			\end{prop}
			
			\begin{pf}
				The proof follows from the definitions of a discriminatory input and the performance index, and is hence omitted.
			\end{pf}
			
			\begin{remark}
				Note that numerical algorithms typically return a sub-optimal solution $u^*$ to the optimization problem \eqref{eq:ID_opt4}. To prove feasibilty of fault diagnosis, we need $\gamma(u)>0$ for at least one $u \in \mathcal{U}$. Recall that the optimization problem \eqref{eq:ID_opt4} exclusively involves positive semi-definite quadratic expressions. Thus, we know that $\gamma(u) \ge 0,\, \forall u \in \mathcal{U}$. Hence, to prove feasibility we require for at least one of the possible locally optimal solutions $u^*$, $\gamma(u^*)$ is non-zero.
			\end{remark}
%
%
\section{Robust fault diagnosis}
\label{sec:robust}
In this section, a strategy is proposed for robust fault diagnosis. Sufficient conditions on the maximal size of perturbations $\Delta_u \in \mathbf{\Delta}_u$ are derived, such that robust fault diagnosis is guaranteed. This leads us to the design of a robust discriminatory input.
%
%
	\subsection{Fault diagnosis - robust analysis}
	\label{sec:FDrobustCase}
\begin{figure}[!t]
		\centering
		\includegraphics[scale = 0.4]{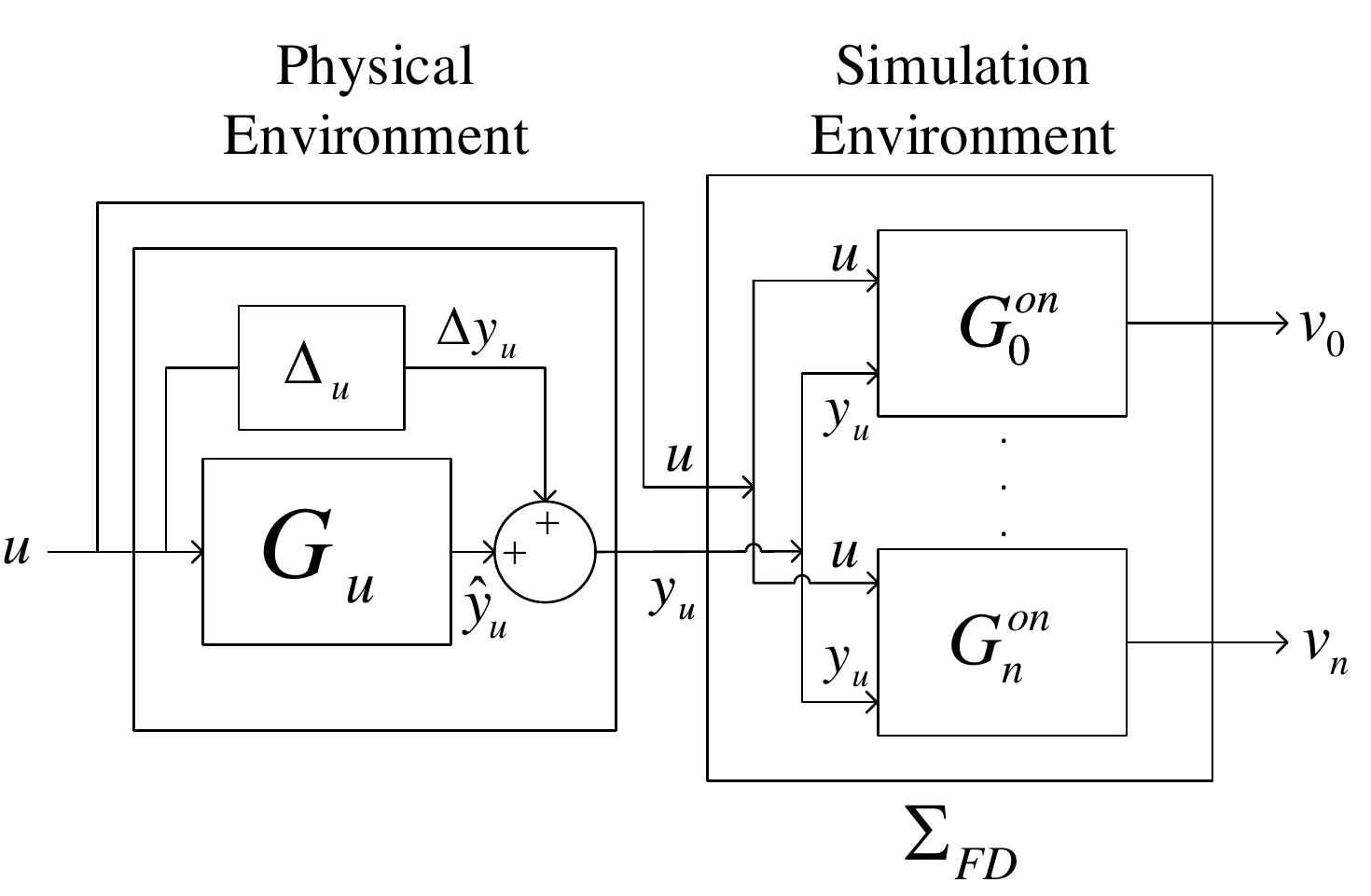}
		\caption{System interconnection $\Sigma_{FD}$ for fault diagnosis - robust case.}
		\label{fig:robFaultDiagnosisIc}
	\end{figure}

	Consider the system interconnection shown in Fig. \ref{fig:robFaultDiagnosisIc}. The definition of the output-nulling representations remains the same as in \eqref{eq:ONrep}, i.e. they are defined for the systems ${G_i \in \mathcal{S}_G}$. $\Sigma_{FD}$ is the parallel connection of the output nulling representations $G_i^{on}$, as shown in Figure \ref{fig:robFaultDiagnosisIc}. These output-nulling representations are assumed to be normalized with respect to the finite-horizon $\ell_2$ induced norm on the interval $[0,T_+]$ (see Appendix \ref{app:normalizeON}).
	
	Let $i^*$ be the index of the true system. It can be easily seen that, for any random realization of the uncertainty $\Delta_u \in \mathbf{\Delta}_u$, the residual signal $v_{i^*}$ corresponding to the true system $G_{i^*}$, initialized by the initial condition $x^*_{{i^*},0}$, can no longer be guaranteed to be identically 0. 
	
	For a given $u \in \mathcal{U}$, let 
	\begin{equation}
		y_u = G_{u,\Delta} u = (G_u + \Delta_u)u = \hat{y}_u + \Delta y_u
		\label{eq:yuDecomposition}
	\end{equation} be the response of the unknown system. Feeding the trajectory $w = \textrm{col}(u, y_u)$ to the output-nulling representation $G_j^{on}$ for some $j \in \mathcal{N}_G$, we obtain the residual:
	\begin{align}
		v_{j} &= G_j^{on} \begin{pmatrix}
u\\ 
\hat{y}_u + \Delta y_u
\end{pmatrix} \nonumber\\
		& = G_j^{on} \left( \begin{pmatrix}
u\\ 
\hat{y}_u
\end{pmatrix} + \begin{pmatrix}
0\\ 
\Delta y_u
\end{pmatrix} \right) \nonumber \\
		& = \hat{v}_j + \Delta v_j, \label{eq:robFDcase1}
	\end{align}
	where, $\hat{v}_j$ is the residual generated due to the deterministic system $G_u$ and $\Delta v_j$ is the residual generated due to the additive uncertainty $\Delta_u$. 
	The linear decomposition of the residual signals leads us to the following result for fault diagnosis in the robust case:
	
	\begin{thm}
		\label{th:FDrobust}
		Let $u^*$ be the optimal discriminatory input defined in Theorem \ref{th:optInputSG}. If 
		\begin{equation}
			\underset{\mathbf{\Delta}_u}{\max} \| \Delta_u \|_H < \gamma(u^*),
			\label{eq:robustInequality}
		\end{equation}
		then, 
		\begin{enumerate}[i)]
			\item there exists an input $u \in \mathcal{U}$ that guarantees robust fault diagnosis for all $\Delta_u \in \mathbf{\Delta}_u$.
			\item If a fault diagnosis experiment is conducted with input $u$ satisfying \eqref{eq:robustInequality}, then
			\begin{equation}
			j^* = \underset{j}{\arg \min} \|v_j\|_{[0,T_+]}
			\end{equation}	
			correctly diagnoses the uncertain system $G_{u,\Delta}$, in the sense that
			\begin{equation}
				G_{u,\Delta} = G_u + \Delta_u \in \mathcal{G}_{j^*}.
			\end{equation}
		\end{enumerate}
	\end{thm}
	
	\begin{pf}
	If $j = i^*$, from \eqref{eq:robFDcase1} we get $\hat{v}_j = 0$ by definition of output-nulling representations. If $j \neq i^*$, by definition of the performance index $\gamma(u)$ we get $\|\hat{v}_j\| \ge \gamma(u)$ for any $u \in \mathcal{U}$. In Section \ref{sec:IDdeterministic}, the input $u^*$ that maximizes $\gamma(u)$ was computed. Thus, from \eqref{eq:robFDcase1} we get that if $\|\Delta v_j \| < \gamma(u^*)$ for some $u \in \mathcal{U}$, then:
	\begin{itemize}
		\item for $j=i^*$, we get $\|v_j\| < \gamma(u^*)$, and
		\item for $j \ne i^*$, we get $\|v_j\| \geq \gamma(u^*)$.
	\end{itemize}
	The largest energy in the residual $\Delta v_j$ for any sample $\Delta_u \in \mathbf{\Delta}_u$ is given by $\|\Delta_u\|_H$. Thus, if 
	\begin{equation}
		\| \Delta_u \|_H < \gamma(u^*), \; \forall \Delta_u \in \mathbf{\Delta}_u \label{eq:FDrobust_sufficient}
	\end{equation}
	then at least $u^*$ achieves robust fault diagnosis. This proves statement i). It also follows that for $j=i^*$, $\|v_j\|$ is minimal for any $u$ obtained from statement i). This concludes the proof. 
	\end{pf}
	
	Theorem \ref{th:FDrobust} provides the solution to RFDP. It can be algorithmically implemented as follows.
	
	\begin{enumerate}[Step 1.]
		\item Conduct the fault diagnosis experiment with a discriminatory input $u(k),\; k \in [-T_-,0)$ on the unknown system $G_{u,\Delta}$ (see Figure \ref{fig:robFaultDiagnosisIc}).
		\item Simulate the system $\Sigma_{FD}$ with trajectory $w = \textrm{col}(u,y_u)$ and initial state as in Appendix \ref{app:initState}.
		\item Compute the $\ell_2$ norm of each of the residual signals $v_j(k),\; k \in [0,T_+]$. Let $j^* = \arg \underset{j}{\min} \|v_j\|_{[0,T_+]}$.
		\item $G_{u,\Delta} \in \mathcal{G}_{j^*}$ is the proposed diagnosis.
	\end{enumerate}
	
	All operations involved in this procedure are algebraic or can be done in polynomial time. Hence, a fault can be robustly diagnosed in polynomial time. 
		

\subsection{Robust input design}
\label{sec:robustID}
\begin{figure} [!t]
				\centering
				\includegraphics[scale = 0.35]{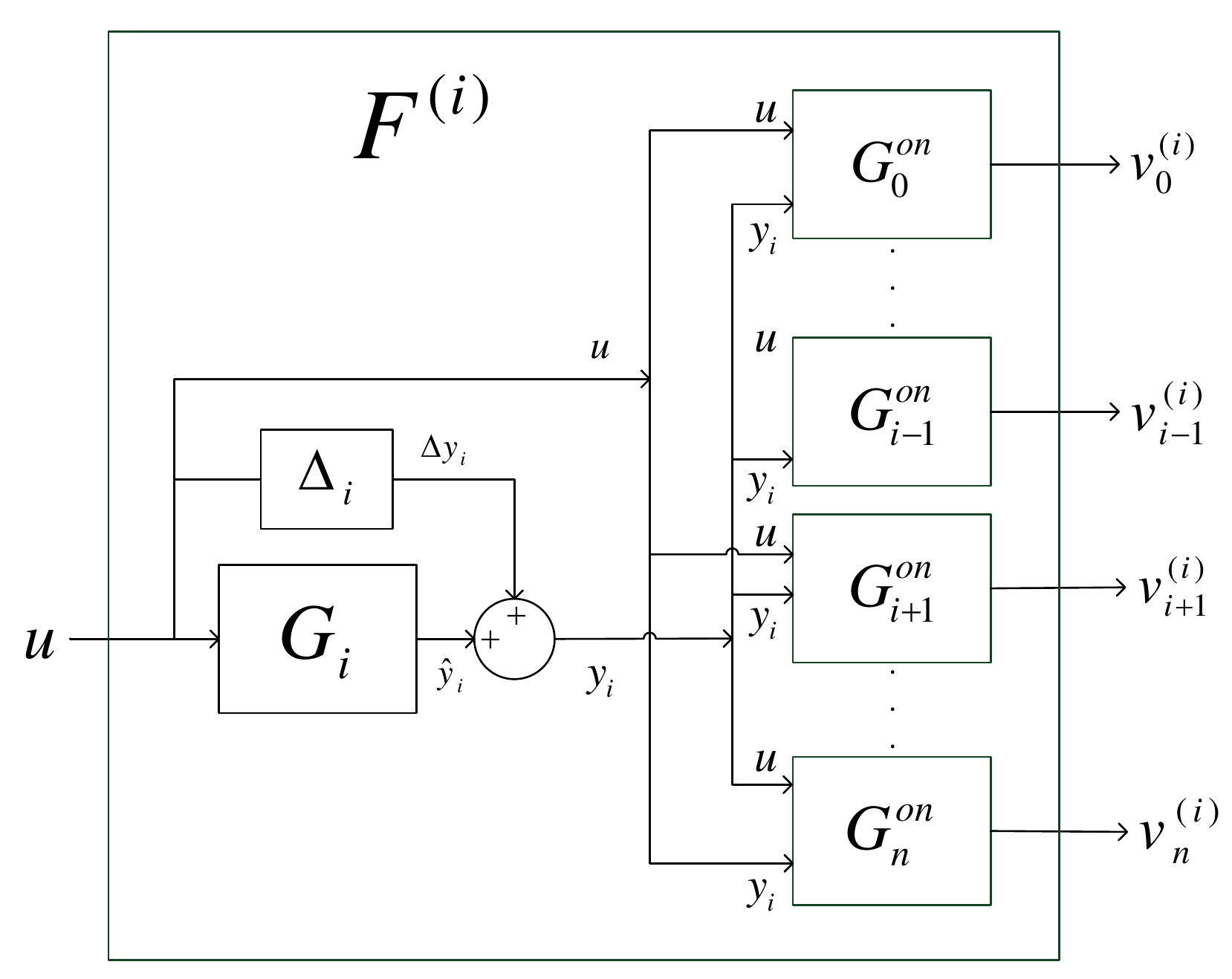}
				\caption{Modified system interconnection $F^{(i)}$ for a fixed $i$.}
				\label{fig:inputDesignFrameworkrob}
	\end{figure}

	In this section, we analyse the problem of input design for robust fault diagnosis. All definitions and propositions in this section follow directly from the proof of Theorem \ref{th:FDrobust}. Consider the modified system interconnection $F^{(i)}$ for some $i \in \mathcal{N}_G$, shown in Fig \ref{fig:inputDesignFrameworkrob}. Let
	\begin{equation*}
		y_i = (G_i + \Delta_i)u = \hat{y}_i + \Delta y_i,
	\end{equation*}
	and 
	\begin{equation*}
		v_j^{(i)} = \left( G_j^{on} \begin{pmatrix}
I \\
G_i + \Delta_i
\end{pmatrix} \right) u = \hat{v}_j + \Delta v_j; \; \forall j\in \mathcal{N}_G^{\bar{(i)}}.
	\end{equation*}
	
	\begin{definition}
		An input sequence $u \in \mathcal{U}$ is called \textit{robustly discriminatory} with respect to system $G_i$ and uncertainty $\mathbf{\Delta}_i$ if it satisfies
		\begin{equation}
			\label{eq:IDrobust_disc_Gi}
			\| \hat{v}_j^{(i)}\|_{[0,T_+]} \ge \| \Delta y_i \|_{[0,T_+]},	\, \forall \, j\in \mathcal{N}_G^{\bar{(i)}}, \forall \Delta_i \in \mathbf{\Delta}_i.
		\end{equation}
	\end{definition}
	Recall the performance index $\gamma^{(i)} (u)$ defined in Equation \eqref{eq:ID_Fi_lb1}, and the corresponding optimal input $u^{(i)*}$ that maximizes the performance index. Define $\mathcal{U}_{\mathrm{rob}}^{(\bar{i})}$ as the set of all robust discriminatory inputs with respect to the system $G_i$.
	
	\begin{prop}
		If the sufficient condition for robust fault diagnosis :
		\begin{equation*}
			\|\Delta_i\|_H \le \gamma(u^{(i)*})
		\end{equation*}
		 holds $\forall \Delta_i \in \mathbf{\Delta}_i$, then the set $\mathcal{U}_{\mathrm{rob}}^{(\bar{i})}$ is non-empty, and the input $u^{(i)*}$ obtained from \eqref{eq:ID_opt2} is also the \emph{optimal robustly discriminatory input} with respect to $G_i$ and $\mathbf{\Delta}_i$.
	\end{prop}
	
	The proposition is easy to verify since the input $u^{(i)*}$ maximizes the energy threshold $\gamma^{(i)}(u)$ between the signals $\Delta y_i$ and $\hat{v}_j^{(i)},\; \forall j \in \mathcal{N}_G^{(\bar{i})}$. Thus, the performance index $\gamma^{(i)}(u)$ can also be interpreted as an energy threshold between the residual signal corresponding to the true diagnosis and all other residual signals (in Figure \ref{fig:inputDesignFrameworkrob}).
	
	Now, consider the system interconnection $F$ as shown in Fig. \ref{fig:inputDesignFramework2} comprising the modified system interconnections $F^{(i)}$. 
	
	\begin{definition}
		An input sequence $u \in \mathcal{U}$ is called \textit{robustly discriminatory} with respect to the set of systems $\mathcal{S}_G$ if it satisfies
		\begin{multline}
			\label{eq:IDrobust_disc_full}
			\|\hat{v}_l\| \ge \| \Delta y_i\|, \; \forall \, i \in \mathcal{N}_G, \, \forall l\in [ni+1,n(i+1)],\\
			\quad \forall \Delta_i \in \mathbf{\Delta}_i
		\end{multline}
	\end{definition}
	Recall the performance index $\gamma(u)$ defined in Equation \eqref{eq:IDexact_gam_full} and the optimal input $u^*$ that maximizes the performance index. Define the set $\mathcal{U}_{\mathrm{rob}}$ as the set of all robustly discriminatory inputs $u \in \mathcal{U}$ with respect to all systems in $\mathcal{S}_G$.
	
	\begin{prop}
		If the sufficient condition for robust fault diagnosis 
		\begin{equation*}
			\|\Delta_i\|_H \le \gamma(u^*)
		\end{equation*}
		holds for all $\Delta_i \in \mathbf{\Delta}_i,\, i \in \mathcal{N}_G$, then the set $\mathcal{U}_{\mathrm{rob}}$ is non-empty, and the input $u^*$ obtained from Equations \eqref{eq:ustar} is also the \emph{optimal robustly discriminatory input} for the set of systems $\mathcal{S}_G$.
	\end{prop}
	
	Again, the proposition can be verified by noting that the input $u^*$ maximizes the energy threshold $\gamma(u)$ between the signals $\Delta y_i, i\in \mathcal{N}_G$ and the corresponding residuals $\hat{v}_l,\; \forall l \in [ni +1,n(i+1)]$. Thus, $u^*$ is the proposed robustly discriminatory input.
%
%
\section{Numerical simulation}
\label{sec:example}
We consider four systems - one nominal system $G_0$, and three faulty systems $\mathcal{G}_1$, $\mathcal{G}_2$ and $\mathcal{G}_3$. In this section, the systems considered for the input design and fault diagnosis problem are described. This will be followed by the optimal (robust) discriminatory input design. Subsequently, the designed input is used to diagnose faults in the system.

The systems are modelled as $4^{th}/6^{th}$ order systems with variations and uncertainties in parameters. The systems are described by the following parametric transfer function:
\begin{multline}
	G_i(q) = g \left( \frac{q^{-2} + b_1 q^{-1} + b_2}{q^{-2} + a_1 q^{-1} + a_2} \right) \left( \frac{q^{-2} + b_3 q^{-1} + b_4}{q^{-2} + a_3 q^{-1} + a_4} \right) \\
	\quad \left( \frac{q^{-2} + b_5 q^{-1} + b_6}{q^{-2} + a_5 q^{-1} + a_6} \right)
\end{multline}
The parameters corresponding to each of the systems is given in Table  \ref{tab:modelParameters}.

\begin{figure} [!t]
		\centering
		\includegraphics[width = 2.5in]{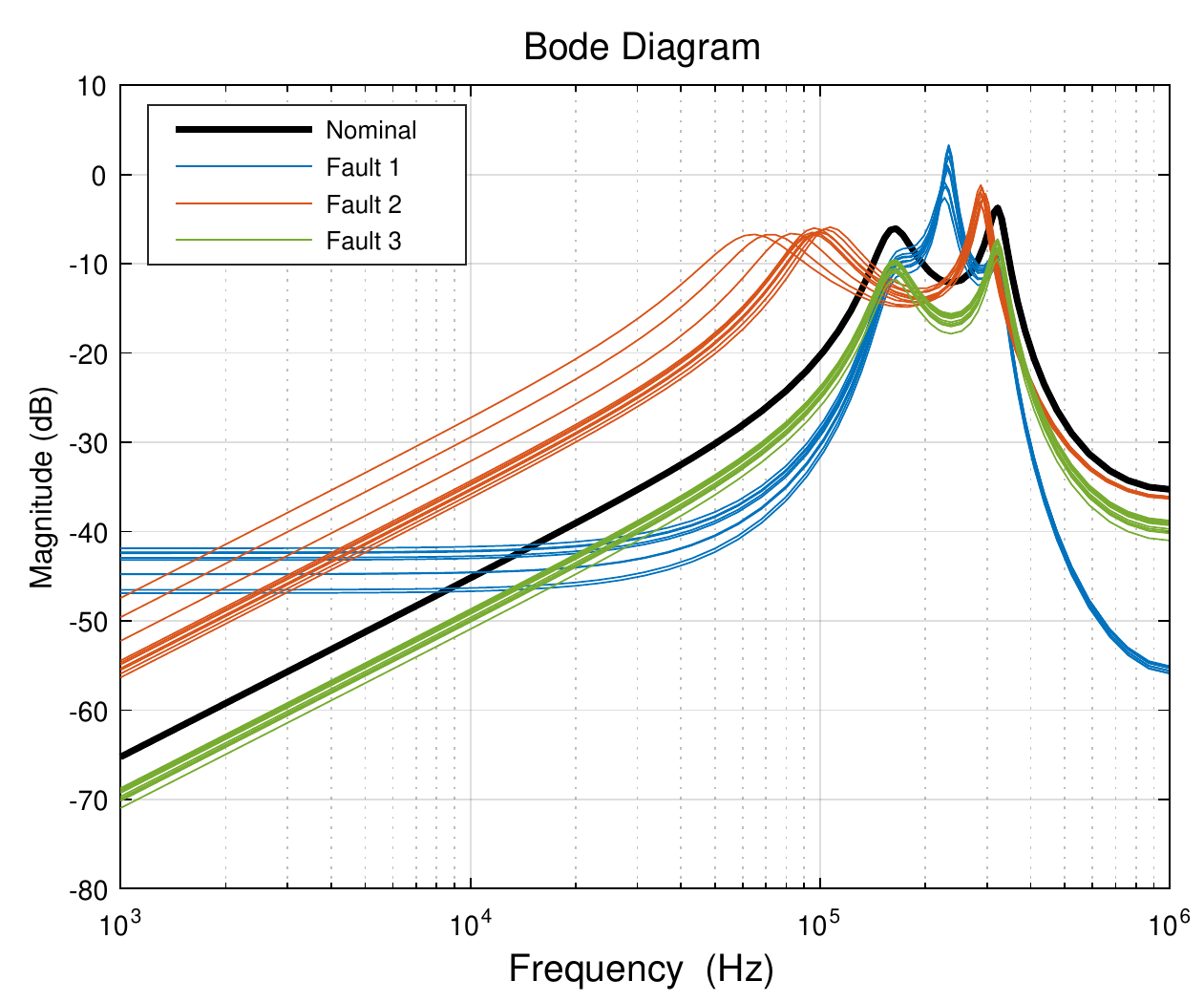}
		\caption{Bode magnitude plots of the systems.}
		\label{fig:sysBode}
\end{figure}

\begin{table}[!t]
{\tiny
\centering
\renewcommand{\arraystretch}{1.3}
\caption{Model parameters}
\label{tab:modelParameters}
\begin{tabular}{|c|c|c|c|c|}
\hline
Param. & $G_0$ & $G_1$ - Fault 1 & $G_2$ - Fault 2 & $G_3$ - Fault 3 
   \\ \hline \hline
g         & -0.0074 & -0.0074           & -0.0074             & $-0.0037 \pm 15 \%$ \\
$a_1$     & -1.6840 & -1.6840           & $-1.8524 \pm 2\%$   & -1.6840 \\
$a_2$     & 0.8839  & 0.8839            & 0.8839              & 0.8839  \\
$a_3$     & -1.0040 & -1.0040           & $-1.1646 \pm 2\%$   & -1.0040 \\
$a_4$     & 0.8971  & 0.8971            & $0.9419 \pm 1.5 \%$ & 0.8971  \\
$a_5$     & 0       & -1.45             & 0                   & 0       \\
$a_6$     & 0       & $0.9345\pm 2\%$ & 0                   & 0       \\
$b_1$     & -1.2194 & -1.2194           & -1.2194             & -1.2194 \\
$b_2$     & 0.2194  & 0.0022            & 0.2194              & 0.2194  \\
$b_3$     & -1.7170 & -1.7170           & -1.7170             & -1.7170 \\
$b_4$     & 7.0670  & 7.0670            & 7.0670              & 7.0670  \\
$b_5$     & 0       & -15               & 0                   & 0       \\
$b_6$     & 0       & $20 \pm 10 \%$    & 0                   & 0       \\ \hline
\end{tabular}
}
\end{table}
The bode magnitude plots of random samples of the chosen systems is shown in Fig. \ref{fig:sysBode}. These systems were chosen as they cover a rich set of variations, as can be seen from the figure. $\mathcal{G}_1$ exhibits an extra mode compared to the nominal system, $\mathcal{G}_2$ exhibits variations in the existing modes of the nominal system and $\mathcal{G}_3$ is merely a damped version of the nominal system.

	The input-output windows are chosen to be $T_- = 32$ and $T_+ = 32$ samples respectively. The sampling frequencies of each of the systems is 2 MHz.
%
	%
	\subsection{Input Design}
	The input design block $F$ is constructed. This system is normalized with respect to the finite horizon Hankel norm, as shown in Appendix \ref{app:normalizeON}. The Matlab routine \textit{fmincon()} was used to solve the non-linear optimization problem in Equation \eqref{eq:ID_opt}. This yields a locally optimal state $\zeta(0)^*$ and a corresponding locally optimal input $u^*$. The optimal input is shown in Fig. \ref{fig:FD_uncertain_complete}. The performance index (defined in \eqref{eq:IDexact_gam_full}) achieved for the exact case is $\gamma(u^*) = 0.0812$.
	
	
	\subsection{Robust Fault Diagnosis}
\begin{figure} [!t]
		\centering
		\includegraphics[width = 0.95\linewidth]{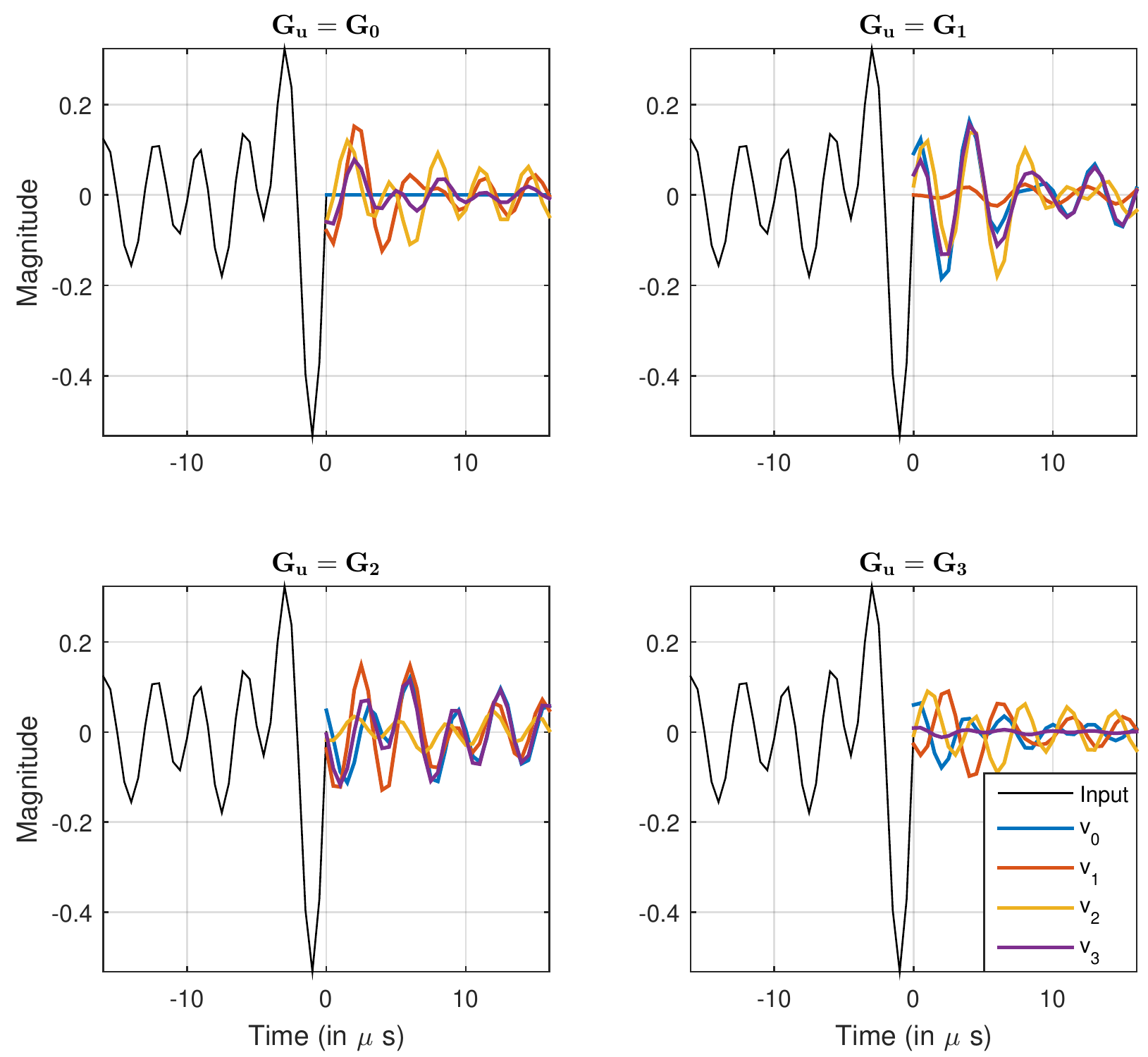}
		\caption{Residual signals from simulation experiments for the robust case.}
		\label{fig:FD_uncertain_complete}
	\end{figure}

	For these simulations, the unknown system $G_{u,\Delta}$ is chosen as the worst case system $G_{i,\Delta}$ for each $i \in \mathcal{N}_G$. Here, worst-case implies that the sample of uncertainty $\Delta_i$ maximizes the norm $\|\Delta_i\|_{\infty}, \, \forall \Delta_i \in \mathbf{\Delta}_i$.  The $\ell_2$ norms of the residual signals produced in each of the four simulations is presented in Table  \ref{tab:FDsim_robust}. For each experiment, the minimal norm, and thereby the proposed diagnosis, is underscored. Note that for experiment 1, the norm $\|v_0\|=0$, since the system $G_0$ is modelled without uncertainties. For all other experiments, the minimal norm is non-zero, due to the modelled uncertainties. Nevertheless, the proposed diagnosis for each experiment corresponds with the true diagnosis. This is in line with the strategy proposed in Section \ref{sec:FDrobustCase}. The residual signals obtained from the experiments are shown in Fig. \ref{fig:FD_uncertain_complete}.

	\begin{table}[!t]
		\vspace*{0.08in}
		\centering
		\caption{Results from fault diagnosis simulations for the robust case.}
		\label{tab:FDsim_robust}
		\begin{tabular}{|c|c|c|c|c|c|}
		\hline
		\begin{tabular}[c]{@{}c@{}}Experiment\\ Number\end{tabular} & \begin{tabular}[c]{@{}c@{}}Experiment\\ Assumption\end{tabular} & $\|v_0\|$ & $\|v_1\|$ & $\|v_2\|$ & $\|v_3\|$ \\ \hline
		1                                                           & $G_u = G_0$                                                     & {\ul 0}   	& 0.3539    		& 0.3468    		& 0.1647    \\ 
		2                                                           & $G_u = G_{1,\Delta}$                                                     & 0.4604    	& {\ul 0.0803}   	& 0.4886    		& 0.4229    \\
		3                                                           & $G_u = G_{2,\Delta}$                                                     & 0.3892    	& 0.4877    		& {\ul 0.1518}   	& 0.4088    \\
		4                                                           & $G_u = G_{3,\Delta}$                                                     & 0.1690    	& 0.2640   		& 0.2803    		& {\ul 0.0250}    \\ \hline
		\end{tabular}
	\end{table}
	
%
%
\section{Conclusions and recommendations}
\label{sec:conclusions}
In this paper, we proposed a new methodology for fault diagnosis and auxiliary input design for discrete-time stable LTI systems with additive LTI perturbations and finite time horizon restrictions. For the input design problem, a performance index $\gamma(u)$ is introduced, which relates to the ability of the input to diagnose the system. Numerical simulation results were presented to illustrate the two methods.

This work can further benefit from experimental validation on a test setup. A natural extension of this work would be the addition of measurement noise to the system description. Another direction for extension of the present work would be to consider different classes of uncertainties such as multiplicative or structured uncertainty.
~	
\appendix
~
\section{Initialization of output-nulling representations}
\label{app:initState}
Consider the output-nulling representation $G_i^{on}$ of system $G_i$ given in Equation \eqref{eq:ONrep}, where $k \in [0,T_+]$, input $u \in \mathcal{U}$ and output $y_i \in \mathcal{Y}$. Since $u(k) = 0 \ \forall k\in[0,T_+]$, the trajectory $w$, defined on the interval$[0,T_+]$, associated with the system $G_i$ is given by $w(k) = \begin{pmatrix}
0\\ 
y(k)
\end{pmatrix}$. The residual signal $v_i(k)$ over the horizon $[0,T_+]$ can now be expressed in terms of the initial state $x_i(0):=x_{i,0}$ of the output-nulling system $G_i^{on}$, as follows:
{\small
\begin{multline}		
	\begin{pmatrix}
	v_i(0)\\ 
	v_i(1)\\ 
	\vdots\\ 
	v_i(T_+)
	\end{pmatrix} = \underset{M_i}{\underbrace{\begin{pmatrix}
	\mathcal{C}_i\\ 
	\mathcal{C}_i\mathcal{A}_i\\ 
	\vdots\\ 
	\mathcal{C}_i\mathcal{A}_i^{T_+}
	\end{pmatrix}}} x_{i,0} \\ \quad
	\underset{N_i}{\underbrace{\begin{pmatrix}
	\mathcal{D}_i & 0 & 0 & \cdots  & 0\\
	\mathcal{C}_i\mathcal{B}_i & \mathcal{D}_i & 0 & \cdots & 0\\ 
	\mathcal{C}_i\mathcal{A}_i\mathcal{B}_i & \mathcal{C}_i\mathcal{B}_i & \mathcal{D}_i & \cdots & 0\\ 
	\vdots  & \vdots & \vdots & \ddots  & \vdots\\ 
	\mathcal{C}_i\mathcal{A}_i^{T_+-1}\mathcal{B}_i & \mathcal{C}_i\mathcal{A}_i^{T_+-2}\mathcal{B}_i & \mathcal{C}_i\mathcal{A}_i^{T_+-3}\mathcal{B}_i &  & \mathcal{D}_i
	\end{pmatrix}}} \begin{pmatrix}
w(0)\\ 
w(1)\\ 
\vdots\\ 
w(T_+)
\end{pmatrix} \label{eq:initState}
\end{multline}}
The problem of finding an initial state $x_{i,0}$ that produces the smallest residual signal $v_i$, measured in the $\ell_2$ norm, given a trajectory $w$, can be framed as the following least squares problem:
\begin{equation}
	x_{i,0}^* = \underset{x_{i,0} \neq 0}{\arg \min} \; \sum_{k=0}^{T_+} \|v_i(k)\|^2, \label{eq:initStateLS}
\end{equation}
where $v_i$ satisfies \eqref{eq:initState} for a given trajectory $w$. This problem can be algebraically solved, and the solution is given by:
\begin{equation}
	x_{i,0}^* = -(M_i^T M_i)^{-1}M_i^T N_i w,
\end{equation}
where the matrices $M_i$ and $N_i$ are defined in \eqref{eq:initState}. A given trajectory $w$ can be attributed to the system $G_i$ if and only if the residual produced by the system $G_i^{on}$, with initial state $x_{i,0}^*$, produces a residual $v_i$ that satisfies $v_i(k) = 0 \ \forall k \in [0,T_+]$.

Since $v_i$ is defined on the horizon $[0,T_+]$, the least squares problem described in Equation \eqref{eq:initStateLS} does not take into account the input applied in the past. This may lead to a situation in which a trajectory $w$ may be attributed to more than one system. For instance, consider two systems $G_1$ and $G_2$ that differ by a scalar factor $k (\neq 0 \text{ or } 1)$, i.e. $G_1 = k G_2$. If the past input is not taken into consideration, any transient output obtained from system $G_1$ can also be obtained from system $G_2$ by choosing the appropriate initial condition, which will be the solution of the least squares problem. 

To explicitly incorporate the information from the past inputs, the following work-around can be used. Consider the state space representation of the system $G_i$. Simulate $G_i$ with the past input $u(k), \, \forall k \in [-T_-,0)$. The final state vector $x_i(0)$ can be used to initialize the system $G_i^{on}$. This is true since there always exists a state-space representation $G_i^{on}$ of the output nulling system, with the same state vector $x_i$ as the original system $G_i$.
%
%
\section{Normalization of output-nulling representation}
\label{app:normalizeON}
An output-nulling representation $G_i^{on}$ can be normalized by simply scaling the output equation in Equation \eqref{eq:ONrep} by a non-singular matrix $R_i$, such that, for a chosen system norm, $\|\tilde{G}_i^{on}\| = 1$, where $\tilde{G}_i^{on}$ is the resulting scaled representation. As explained in Section \ref{sec:detCaseMath}, the $\ell_2$ induced norm, defined in \eqref{eq:l2norm}, and the finite horizon Hankel norm, defined in \eqref{eq:Hankelnorm}, are of particular interest. The normalized system representation $\tilde{G}_i^{on}$ for $i \in \mathcal{N}_G$ is given by:
\begin{subequations}
	\label{eq:normONrep}
	\begin{eqnarray}
		x_i(k+1) &=& \mathcal{A}_i x_i(k) + \mathcal{B}_i w(k), \label{eq:normONrep1}\\
		\tilde{v}_i(k) &=& R_i \mathcal{C}_i x_i(k) + R_i \mathcal{D}_i w(k), \label{eq:normONrep2}\\
	&&\quad x_i(k) \in \mathbb{R}^{n_i}.\nonumber
	\end{eqnarray}
\end{subequations}
It must be noted that normalization changes the transfer function of $G_i^{on}$ in the sense that the transfer function $G_i^{on} \ne \tilde{G}_i^{on}$. However, $G_i^{on}$ and $\tilde{G}_i^{on}$ represent the same model in the sense that for any $w$ in \eqref{eq:ONrep} and \eqref{eq:normONrep}, we have that 
\begin{equation}
	v_i = 0 \Leftrightarrow \tilde{v}_i = 0.
\end{equation}
The fault diagnosis methodology necessitates the comparison of the residual signals $v_i$ generated by the trajectory $w = \textrm{col}(u,y_i)$, both of which are defined over the future horizon $[0,T_+]$. Thus, the output-nulling representations must be normalized with respect to the $\ell_2$ induced norm, defined in \eqref{eq:l2norm}. On the other hand, the input design methodology requires the comparison of the residuals $v_i$ generated by the input $u$, the latter being defined on the past interval $[-T_-,0)$. Thus, in the case of input design, the output-nulling representation must be normalized with respect to the finite horizon Hankel norm given in \eqref{eq:Hankelnorm}.

In either case, the scaling factor $R_i$ can be found using a bisection algorithm over a feasible range of scaling factors, until the condition $\|\tilde{G}_i^{on}\| = 1$ is satisfied.

Note that, in general, an output-nulling system is normalized by ``output-injection'' and ``output-scaling'', resulting in a co-inner representation $G$ such that $GG^* = I$, where $G^*$ is the adjoint of the system. This implies that $G$ is normalized with respect to the $H_{\infty}$ norm. The method for obtaining co-inner representations is elaborated in \cite{weiland1991theory}.
%
%
{\footnotesize
\bibliography{IFAC_WC_2017_bib}}             
                                                   







\end{document}